\documentclass[twocolumn,aps,pra,groupedaddress,amsfonts, 10pt]{revtex4-1}

\usepackage{amsmath}
\usepackage{amssymb}
\usepackage{graphicx}
\usepackage{bm}
\usepackage{array}
\usepackage{dcolumn}
\usepackage{hepparticles}
\usepackage{heppennames}
\usepackage{hepnicenames}
\usepackage{epstopdf}
\usepackage{tikz}
\usepackage{xcolor}
\usepackage{multirow}

\newcommand{\ket}[1]{{| {#1} \rangle}}

\definecolor{bostonuniversityred}{rgb}{0.8, 0.0, 0.0}

\begin{document}

\title{A High Motional Frequency Ion Trapping Regime for Quantum Information Science}

\author{A.J. Rasmusson}
\affiliation{Brigham Young University Department of Physics and Astronomy, Provo, Utah 84602, USA}

\date{\today}

\begin{abstract}
We investigate high frequency motional states of trapped atomic ions. Trapped ions in rf traps are confined by an approximate harmonic potential and exhibit quantum motional states that mediate essential techniques in quantum computing, simulation, networking, and precision measurement. However, motional state decoherence mechanisms, heating and dephasing, are broadly limiting: reduced two-qubit gate fidelities; lower fidelity and lifetime of highly nonclassical bosonic states; long laser cooling times; and large recoil heating rates. These also challenge the scalability of increasingly sophisticated protocols. We propose high motional frequency ion trapping as an operating regime that addresses these challenges and reshapes the design landscape for quantum information experiments and quantum control techniques. We report an experimentally motivated investigation of realizing this high-frequency regime and discuss the consequences for laser cooling, motional state coherence, fidelity and lifetime of nonclassical bosonic states, and scalability of experimental runtimes. We report clear design trajectories for ion traps to reach high motional frequency, a new limiting mechanism for laser cooling at these high frequencies, and more than an order-of-magnitude speedup in experimental duty cycles with larger speed ups possible for quantum error correction protocols. Taken together, high motional frequency ion trapping has broad implications for the future of quantum information experiments.
\end{abstract}

\maketitle

\section{Introduction}
\label{sec:introduction}

Trapped atomic ions in a radio-frequency (rf) trap are approximately confined by a harmonic potential. The resulting quantum harmonic oscillator states and Coulomb interactions between co-trapped ions have enabled rich dynamics, diverse interactions, and high-fidelity control of quantum degrees of freedom. For example, in a quantum computing context, motional states enable qubit-qubit interactions necessary for two-qubit gate operations with $>$99.99\% fidelity \cite{hughes2025trapped}, the generation of large entangled states \cite{moses2023race}, and even bosonic quantum error correction protocols \cite{fluhmann2019encoding, matsos2024universal}. Within quantum simulation, motional states mediate effective spin dynamics that emulate a variety of quantum systems \cite{monroe2021programmable} and can directly participate in simulating bosonic degrees of freedom \cite{whitlow2023quantum, than2025observation}. In precision measurement settings, cooling to near the motional ground state reduces atomic clock uncertainty \cite{chen2017sympathetic} and enables quantum logic spectroscopy \cite{schmidt2005spectroscopy}. Additionally, topological phases \cite{niedermeyer2025observation} and quantum synchronization effects \cite{liu2025observation} have been directly observed in the quantum motion of trapped ions.

Across these varied scientific aims, decoherence mechanisms in trapped-ion systems largely acts on motional states and are often limiting. For example, two-qubit gate speeds and fidelities are constrained by the motional frequency and cooling limits respectively \cite{savill2025high, hughes2025trapped}. Experimental runtimes are dominated by slow laser cooling durations \cite{pino2021demonstration, ransford2025helios}. Motional state decoherence rates are anomalously high in rf traps \cite{wineland1998experimental, brownnutt2015ion} leading to motional heating, thermal averaging noise, and second-order Doppler shifts in atomic clocks \cite{rasmusson2021optimized, marshall2025high}. Even the number of ions that can be co-trapped, before structural phase transitions occur, is limited by the motional frequency \cite{d2020radial}. Reaching higher motional frequencies would alleviate these fundamental challenges and enable a new generation of quantum capabilities in trapped-ion systems.

Trapped-ion experiments typically operate near $\sim$ 1-2 MHz with few exceptions. One work indirectly observed a motional frequency of about 50 MHz \cite{jefferts1995coaxial}, but with no follow-up studies near this frequency. Single electrons have been trapped in an rf trap and reached extremely high motional frequencies of 380 MHz \cite{matthiesen2021trapping, mikhailovskii2025trapping}. However, single electrons are still exploratory with, for example, cooling and state readout being an active area of investigation \cite{yu2022feasibility}. In this work, we focus our investigation on the advantages of high motional frequencies of trapped ions which have well-established and high-fidelity techniques necessary for universal quantum control.

In this experimentally motivated work, we systematically investigate advantages of high frequency motional states of trapped ions. Section~\ref{sec:high-secular-frequency} evaluates competing experimental design points for realizing higher motional frequencies. Changes in laser cooling rates and limits are compared between low and high frequency regimes in Sec.~\ref{sec:resolved-cooling}. Decoherence effects on motional states at high frequency are discussed, new cooling limits are introduced, and the fidelity and lifetime of nonclassical bosonic states are investigated in Sec.~\ref{sec:motional-state-coherence}. Finally, in Sec.~\ref{sec:fast-duty-cycles}, the operational runtime of trapped-ion experiments, including quantum error correction protocols, is evaluated for high motional frequency ion traps.

\section{High secular frequency RF-traps}
\label{sec:high-secular-frequency}

\begin{figure*}
    \centering
    \includegraphics[width=0.9\textwidth]{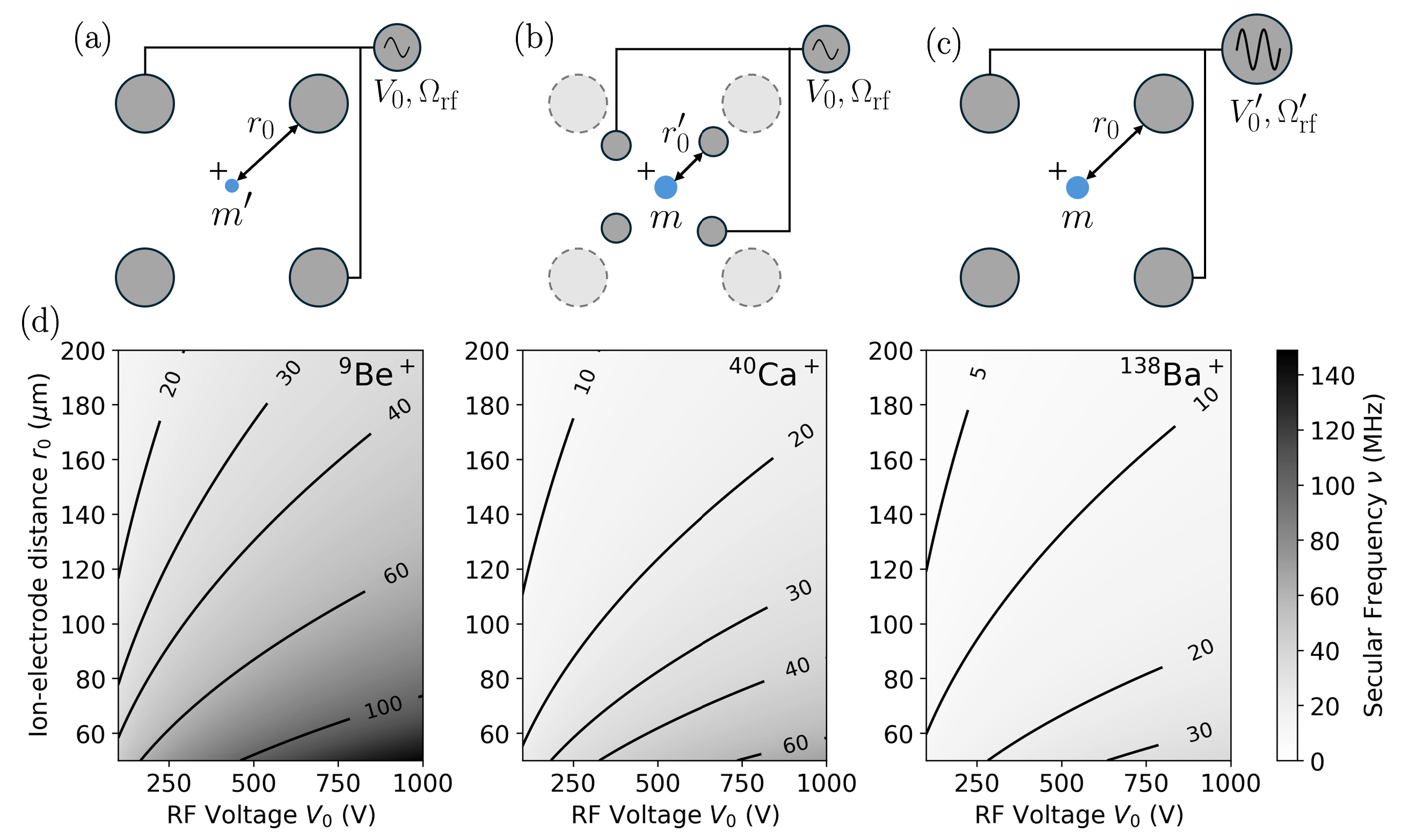}
    \caption{The effect of trap design parameters on confinement frequency $\nu$. The top row of panels are schematic cross-sections of a 4-rod rf trap with labeled experimental variables on each panel representing possible design variations. (a) Reduced ion mass to increase $Q/m$. (b) Reduced ion-electrode distance $r_0$. (c) Increased $V_0$ or $\Omega_{\text{rf}}$. (d) Estimates of the radial confinement frequency $\nu$ for different mass ions, across a range of ion-electrode distances $r_0$ and rf voltage amplitudes $V_0$. Micromotion amplitude is fixed to $q = 0.4$ and contours label the secular frequency $\nu$ at specific values. We assume an in-phase rf drive on one diagonal electrode pair (shown) and an out-of-phase rf drive on the other pair (not shown).}
    \label{fig:trap-params}
\end{figure*}

The rf trap confines charged particles in an effective harmonic oscillator potential using static and oscillating electric fields. For trapped-ion experiments, this potential has largely been on the order of $\sim$1 MHz, but it can be increased by more than an order of magnitude with careful experimental consideration, which regime we now investigate.

In two spatial dimensions, which we refer to generically as the radial dimension, oscillating electric fields $V_0\cos(\Omega_{\text{rf}} t)$ in a quadrupole configuration with ion-electrode distance $r_0$ produce a harmonic pseudopotential under the right experimental conditions. A static electric field confines in the final spatial dimension which is typically kept weaker so ions form a one-dimensional chain.

Now, suppose an atomic ion has a charge-to-mass ratio $Q/m$. The harmonic trajectory along the radial dimension $u(t)$ can be written to first order as
\begin{equation}
    u(t) \approx u_0 \cos(\nu t - \phi)\left(1 + \frac{q}{2}\cos(\Omega_{\text{rf}} t)\right)
\end{equation}
with the secular frequency $\nu$, micromotion amplitude $q$, and initial conditions determining $u_0$ and $\phi$ \cite{wineland1998experimental}. The micromotion amplitude is defined as
\begin{equation}
    q = \frac{2QV_0}{m\Omega_{\text{rf}}^2r_0^2}
\end{equation}
with stable trapping trajectories across values $q = (0, 0.908)$ \cite{wineland1998experimental, wuerker1959electrodynamic}. It quantifies the amplitude of a fast oscillation at frequency $\Omega_{\text{rf}}$ on top of the slower harmonic motion at frequency $\nu$. For small micromotion amplitude $q^2 \ll 1$, micromotion can be neglected and the ion trajectory simplifies to harmonic motion. We will maintain this assumption throughout this work. We leave the exploration of more exotic trajectories for future work.

Assuming strong radial confinement compared to axial confinement, as mentioned previously, the radial harmonic frequency is approximated as
\begin{equation}
    \nu \approx \frac{Q V_0}{\sqrt{2} m r_0^2 \Omega_{\text{rf}}} \;.
\end{equation}

For an order of magnitude increase to 30 MHz or more, there are primarily four experimental design parameters as shown in Fig. \ref{fig:trap-params}. First, the charge to mass ratio $Q/m$. Second, the ion-electrode distance $r_0$. Third, the rf drive frequency $\Omega_{\text{rf}}$. Fourth, the rf drive amplitude $V_0$. We now systematically investigate each option balancing key experimental constraints while maintaining small micromotion amplitude $q^2 \ll 1$.

For singly ionized atoms, a change in the charge-to-mass ratio corresponds to choosing a different atomic species [see Fig. \ref{fig:trap-params}(a)]. Atomic ions with masses ranging from a few atomic mass units $^9$Be$^+$ to nearly 200 $^{174}$Lu$^+$ have been trapped \cite{jefferts1995coaxial, arnold2016observation} implying an order-of-magnitude variation is possible, all other trap parameters being fixed. Higher charged ions have also been trapped, but the often more complex internal state structure and deeper ultraviolet transitions can present significant experimental challenges to key techniques such as laser cooling \cite{kozlov2018highly, dreiling2019capture}. Varying the charge-to-mass ratio $Q/m$ changes both the confinement strength and the micromotion amplitude proportionally, as $\nu \propto Q/m$ and $q \propto Q/m$. Consequently, varying $Q/m$ alone cannot increase $\nu$ without also varying other experimental parameters to maintain the adiabatic condition $q^2 \ll 1$. Interestingly, the Lamb-Dicke parameter $\eta \propto 1/\sqrt{m\nu}$, which quantifies qubit-motion coupling that is essential to many trapped-ion techniques such as two-qubit gates \cite{wineland1998experimental, molmer1999multiparticle}, would remain constant for different atomic species of the same charge. Thus, while strong confinement frequencies could be realized from lighter atomic species, rates for entangling interactions and even quantum logic spectroscopy could remain unchanged.

Varying the ion to electrode distance $r_0$ [see Fig. \ref{fig:trap-params}(b)] changes $\nu$ by $1/r_0^2$, allowing even small reductions in $r_0$ to significantly increase confinement. However, as in the previous case, the micromotion amplitude $q$ has the same scaling $q \propto 1/r_0^2$, so changing $r_0$ alone cannot increase $\nu$ without also varying other experimental parameters to maintain $q^2 \ll 1$. Adjusting the ion–electrode spacing comes with an additional key constraints. The closer a trapped ion is to the electrode surface the larger the coupling to the electrode's noisy electric environment making the ion more sensitive to stray surface charges and enhancing anomalous heating, which scales strongly as $1/r_0^4$ and is already a primary noise source \cite{brownnutt2015ion, sedlacek2018distance}. Technical demands, such as precise geometric tolerances and electrode surface preparation, also increase. Optical access is constrained for smaller traps requiring more tightly focused laser beams to avoid light irradiating electrode surfaces which can produce local surface charging and perturb the trapping potential \cite{sagesser2024}. Smaller trap dimensions increase electrode capacitance which can lead to a reduced Q factor for loaded rf resonators limiting achievable rf voltages and noise filtering \cite{siverns2012application}. Collectively, these effects set challenging technical limits on how far $r_0$ can be reduced while preserving stable and low-noise trapping. Although reducing $r_0$ can significantly increase confinement, the rapidly increasing heating rates and other technical constraints mean that only modest reductions are reasonably achievable. However, ion-electrode distances as small as 30 $\mu$m have been realized, typically in cryogenic systems to mitigate electric field noise \cite{srinivas2021high}.

\begin{figure}
    \includegraphics[width=1.0\linewidth]{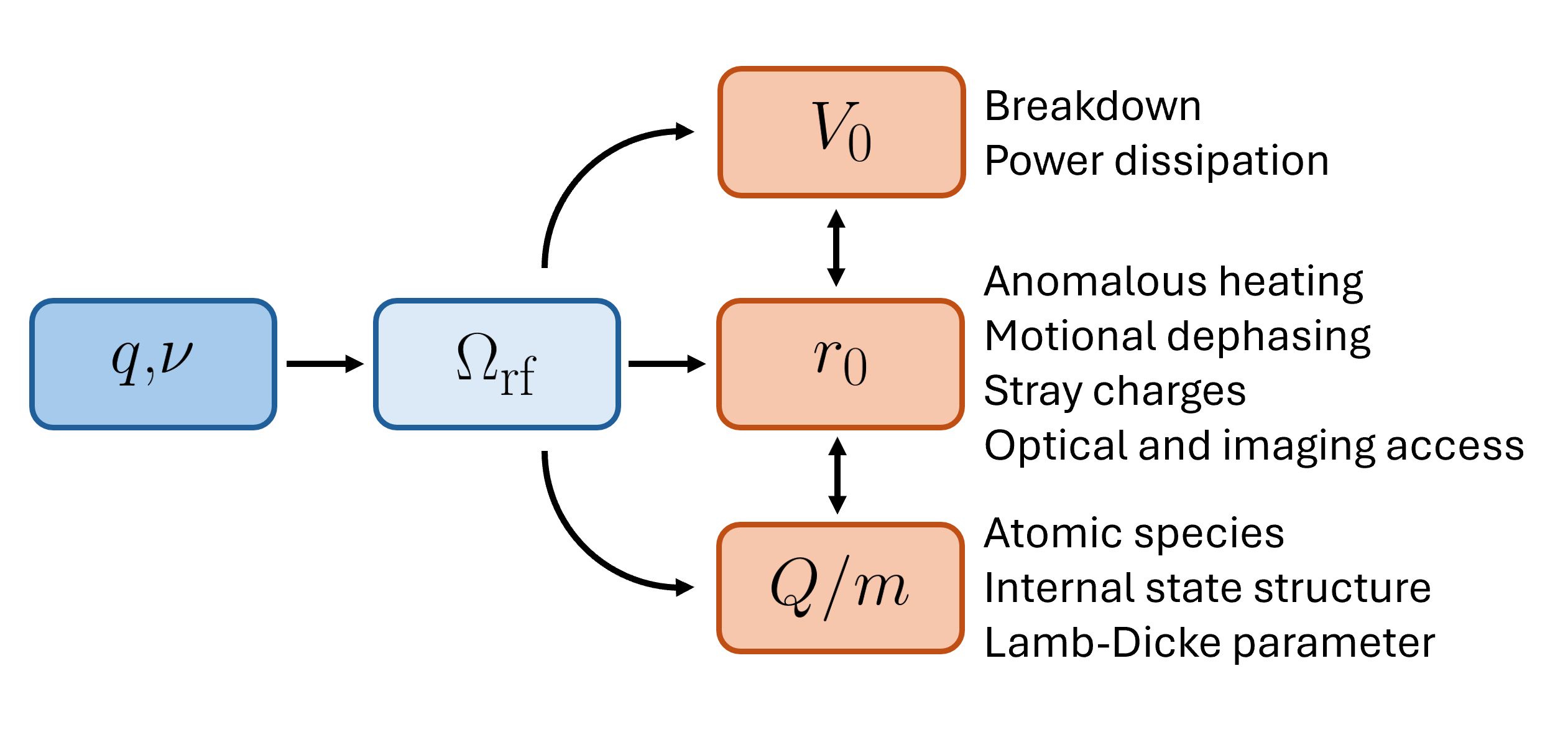}
    \caption{Summary schematic of the design flow for experimentally realizing a high motional frequency ion trap. Starting with the selection of the target micromotion amplitude $q$ and motional frequency $\nu$, $\Omega_{\text{rf}}$ is then fixed by Eq. \eqref{eq:sec-freq-omega-rf}. The remaining experimental design choices, in orange, $V_0$, $r_0$, and $Q/m$ are balanced against each other to meet hardware constraints and practical considerations. Leading experimental design considerations are listed by each variable.}
    \label{fig:trap-design-guide}
\end{figure}

The rf drive amplitude $V_0$ is similar to the previous experimental parameters in that $\nu$ and $q$ have the same scaling $\nu, q \propto V_0$. The rf drive frequency $\Omega_{\text{rf}}$ is the only experimental parameter for which $\nu$ and $q$ exhibit different scaling. While $\nu \propto 1 / \Omega_{\text{rf}}$, the micromotion amplitude scales as $q \propto 1/\Omega_{\text{rf}}^2$. It is natural, therefore, to rewrite $\nu$ in terms of $\Omega_{\text{rf}}$ for fixed $q$
\begin{equation}
    \nu = \frac{q}{2\sqrt{2}}\Omega_{\text{rf}} \;.
\label{eq:sec-freq-omega-rf}
\end{equation}
Thus, for constant $q$, increasing the rf frequency is required to achieve larger $\nu$ (see Fig. \ref{fig:trap-design-guide}). In practice, raising $\Omega_{\text{rf}}$ or $V_0$ introduces technical challenges such as increased ohmic heating and decreased rf resonator Q factors needed to reach large $V_0$ and filter out electrical noise \cite{siverns2012application}. These are primarily technical in nature however and have been overcome in earlier ion trap designs \cite{jefferts1995coaxial}. With the maturity of modern distributed circuit elements, numerical simulation tools, and microfabrication capabilities, rf systems are poised to readily overcome these challenges.

We give two experimentally realistic examples to illustrate the above observations (also see Fig. \ref{fig:trap-design-guide}). First, suppose $^9$Be$^+$ is trapped with the following parameters: $\Omega_{\text{rf}} = 2\pi$ $\times$ 230 MHz, $V_0 = 280$ V, $r_0 = 120$ $\mu$m. The secular frequency is then $\nu = 2\pi$ $\times$ 32 MHz with micromotion amplitude $q = 0.4$---maintaining the adiabatic trapping condition and achieving high secular frequency. Furthermore, and while maintaining $q=0.4$, the voltage $V_0$ and $r_0$ can be balanced against each other to optimize experimental limitations as illustrated in Fig. \ref{fig:trap-design-guide}. For example, if a lower voltage is desired, such as $V_0 = 195$ V, the ion-electrode distance can be reduced $r_0 = 100$ $\mu$m to maintain the secular frequency and micromotion amplitude. Second, to achieve $\nu = 2\pi$ $\times$ 50 MHz a larger rf frequency is key such as $\Omega_{\text{rf}} = 2\pi$ $\times$ 350 MHz. To maintain $q=0.4$, the voltage amplitude and ion-electrode distance can again be balanced with an example choice of $V_0 = 655$ V and $r_0 = 120$ $\mu$m. Pushing further, an ion trap with harmonic frequencies as high as 100 MHz, as shown in Fig. \ref{fig:trap-params}(d), is within reach with a coordinated design of the experimental design choices $\Omega_{\text{rf}}$, $V_0$, $r_0$, and $Q/m$ as summarized in Fig. \ref{fig:trap-design-guide}.

\section{Resolved Doppler cooling rates and limits}
\label{sec:resolved-cooling}

Laser cooling of trapped ions is an essential yet time-consuming step in trapped-ion experiments. Laser cooling can take up more than 60\% of an experimental duty cycle which can lead to idling errors exceeding two-qubit gate errors \cite{pino2021demonstration, yamamoto2025quantum}. Across a wide variety of scientific aims, a trapped-ion experimental cycle typically begins with laser cooling to the Doppler limit---set by spontaneous emission \cite{wineland1979laser, itano1982laser}---followed by sub-Doppler cooling to reach near the motional ground state \cite{rasmusson2021optimized}. These cooling stages ensure that ions remain trapped and are prepared in a well-defined quantum motional state and reduced thermal averaging noise. Operating rf traps in the \textit{resolved} Doppler regime $\nu \gg \Gamma$ \cite{wineland1979laser}, where $\Gamma$ is the natural linewidth of a fast cycling transition ($\Gamma \sim 2\pi \times20$ MHz for many trapped-ion species), the motional sidebands on the optical laser cooling transition become spectrally resolved and the cooling rates and limits fundamentally change compared to the unresolved regime $\nu \ll \Gamma$ as shown in Table \ref{tab:cooling-rate-limits} and Fig. \ref{fig:cooling-rates-and-limits}. The Doppler cooling rate in the resolved regime can be much faster, and the cooling limit can approach the motional ground state eliminating the need for slow and more experimentally complex sub-Doppler cooling techniques.

\subsection{Cooling rates}
\label{sec:cooling-rates}
The laser cooling rate is set by the difference of loss and gain of motional energy from photon absorption and spontaneous emission. Consider a two-level atomic state system interacting with a quantized harmonic oscillator of state $\ket{n}$. In the Lamb-Dicke regime, with Lamb-Dicke parameter $\eta = \sqrt{E_{\text{recoil}} / \hbar \nu}$, first-order processes (i.e. $\ket{n} \rightarrow \ket{n\pm1}$) dominate and the master equation reduces to a rate equation for the motional-state populations $P(n)$ \cite{stenholm1986semiclassical, leibfried2003quantum}:
\begin{align}
    \frac{d}{dt}P(n) &= (n+1)A_-P(n+1) \\
    & - [(n+1)A_+ + nA_-]P(n) + nA_+P(n-1) \nonumber \;,
\end{align}
where $A_{\pm}$ are the transition rates for adding or removing a motional quantum (see Fig. \ref{fig:graph}). These coefficients take the form
\begin{equation}
    A_{\pm} = \eta^2 \Gamma \left[W(\Delta) + W(\Delta \mp \nu)\right]
\end{equation}
with $W(\Delta)$ capturing the lineshape of the driven two-level optical transition with frequency detuning $\Delta$.

The mean motional occupation $\bar{n} \equiv \sum P(n) n$ then evolves according to the rate equation
\begin{equation}
    \frac{d}{dt}\bar{n} = -(A_- - A_+)\bar{n} + A_+
\label{eq:dndt}
\end{equation}
from which the cooling rate is identified as
\begin{equation}
    R_{\text{c}} = (A_- - A_+) \;.
\end{equation}

In the \textit{resolved} Doppler regime, when the cooling laser is tuned near the first-order red sideband ($\Delta=-\nu$), the $A$ coefficients simplify to
\begin{align}
    A_- &\approx \eta^2 R_{\text{sc}} \\
    A_+ &\approx \eta^2R_{\text{sc}}\left(\dfrac{\Gamma}{2\nu}\right)^2
\end{align}
where $R_{\text{sc}} = \Omega^2/\Gamma$ is the on-resonance scattering rate in the low-intensity limit with on-resonance Rabi frequency $\Omega$. The resulting cooling rate becomes
\begin{equation}
    R_{\text{c}} \approx \eta^2 R_{\text{sc}} \;
\end{equation}
to first order in $\Gamma/\nu$. A comparison between this result and the unresolved sideband regime cooling rate is summarized in Table~\ref{tab:cooling-rate-limits}. Cooling occurs when $A_- > A_+$, and in only the resolved sideband regime ($\nu \gg \Gamma$), the heating term $A_+$ is strongly suppressed by the factor $(\Gamma/2\nu)^2$. However, this is balanced against the overall cooling rate itself which scales as $R_{\text{c}} \propto \eta^2 \propto 1/(m\nu)$ leading to seemingly slower cooling rates for increasing secular frequency. However, there is an additional subtlety as the Lamb-Dicke parameter depends on $\nu$ and $m$ in additional to the factor $\nu/\Gamma \ll 1$ which slows the cooling rate for unresolved case.
\begin{table}[t]
    \caption{Comparison of the scaling relations for Doppler cooling rates and limits in the resolved and unresolved sideband regimes with respect to $\nu$. The resolved regime has an additional factor of $\nu/\Gamma \ll 1$ which lowers the cooling rate and increases the cooling limit compared to the unresolved regime.}
    \label{tab:cooling-rate-limits}
    \def\arraystretch{1.5}
    \begin{tabular}{p{2cm}|p{3.3cm}|p{2.8cm}@{}m{0pt}@{}}
    & Unresolved Regime & Resolved Regime \\
    & ($\Gamma \gg \nu$) & ($\nu \gg \Gamma$) & \\
    \hline
     Detuning & $\Delta = -\Gamma/2$ & $\Delta = -\nu$ & \\
    \hline
    Cooling rate & $d\bar{n} /dt \propto -\eta^2 R_{\text{sc}} \nu / \Gamma$ & $d\bar{n}/dt \propto -\eta^2 R_{\text{sc}} $ & \\
    \hline
    Cooling limit & $\bar{n} \propto \Gamma/(2\nu)$  & $\bar{n} \propto [\Gamma/(2\nu)]^2$ & \\
    & ($\bar{n} \gg 1$)  & ($\bar{n} \ll 1$) & \\
    \end{tabular}
\end{table}

For example, when comparing cooling rates across atomic species, the Lamb-Dicke parameter can be identical, but the two different species may be confined by motional frequencies orders-of-magnitude apart. Suppose $^{171}$Yb$^+$ is confined with motional frequency $\nu=2\pi\times 2$ MHz and consequently $\eta \approx 0.06$. A lighter ion such as $^{9}$Be$^+$ can have the same value $\eta\approx 0.06$ but with the order-of-magnitude larger motional frequency $\nu=2\pi\times 50$ MHz. For similar on-resonance scattering rates, the much lighter $^{9}$Be$^+$ ion has a $\times$10 faster cooling rate (see Fig. \ref{fig:cooling-rates-and-limits}). This is a key observation to the subtle dependencies on the experimental design parameters ($Q/m$, $r_0$, $V_0$, $\Omega_{\text{rf}}$) that can change expected scalings when derivations are generalized solely in terms of $\nu$. When motional coupling is relevant, such as with laser cooling, it is more informative to include $m$ and $\nu$ dependencies. With all of these considerations, the resolved Doppler regime can facilitate faster cooling at higher $\nu$.

\subsection{Cooling limits}
\label{sec:cooling-limits}

\begin{figure}[b]
    \centering
    \includegraphics[width=0.98\linewidth]{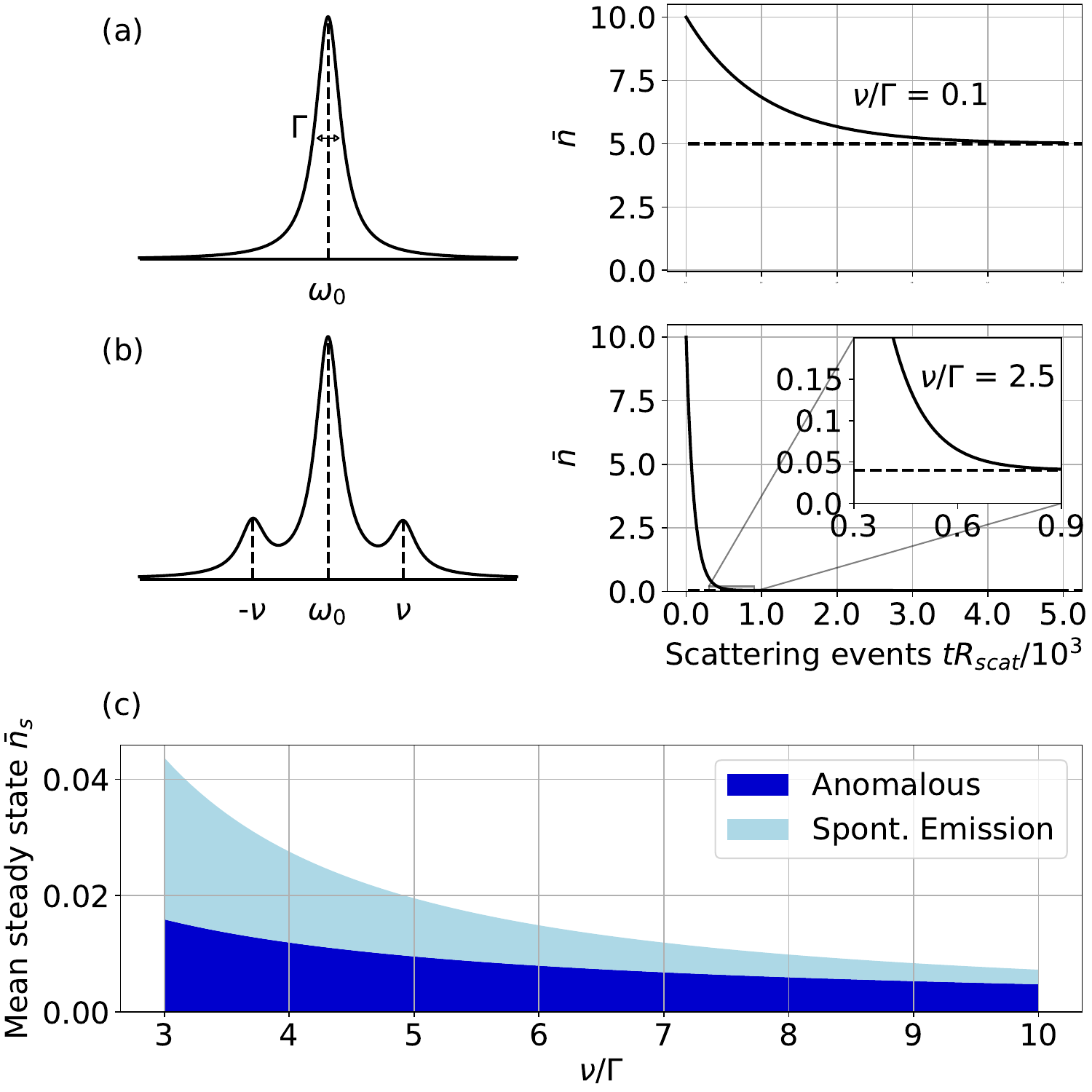}
    \caption{Comparison of Doppler cooling rates and limits between the resolved and unresolved sideband regimes. Considered ratios are $\nu/\Gamma = 0.1$ (a) and $\nu/\Gamma=2.5$ (b) with the same Lamb-Dicke parameter $\eta=0.1$ and initial thermal state $\bar{n}_0=10$. (a) Unresolved regime with atomic linewidth on left and cooling dynamics on the right. (b) Resolved regime with, on the left, the atomic linewidth showing the now spectrally resolved sidebands and, on the right, the cooling dynamics. The resolved regime cools $\sim$10 times faster and to $\sim$100 times lower steady state $\bar{n}$. (c) Contributions to the mean occupation steady state $\bar{n}_s$ in the resolved Doppler regime by spontaneous emission \textit{and} anomalous heating [see Eq. \eqref{eq:steady-state-nbar}].}
    \label{fig:cooling-rates-and-limits}
\end{figure}

The laser cooling limit is set by the balance of the cooling rate and motional excitation rate from spontaneous emission. The steady-state mean occupation number $\bar{n}_s$ is obtained by setting $d\bar{n}/dt = 0$ in Eq.~\eqref{eq:dndt}, giving the cooling limit in the resolved regime
\begin{align}
    \bar{n}_s &= \frac{A_+}{A_- - A_+} \nonumber \\
    & \approx (\Gamma/2\nu)^2 \;.
\end{align}
This steady-state value satisfies $\bar{n} \ll 1$ corresponding to near-ground-state cooling, shown in Fig. \ref{fig:cooling-rates-and-limits}(b), without the need for additional sub-Doppler cooling. This can lead to a sizable reduction in experimental runtime, as laser cooling can take the majority of an experimental cycle, and much lower power laser systems.

A further distinction arises in the motional state distribution after cooling. For resolved Doppler cooling, the final population distributions remains approximately thermal, facilitating direct, accurate, and straight-forward measurement of $\bar{n}$ and simplified numerical propagation of motion related noise \cite{chen2017sympathetic, rasmusson2021optimized}. In contrast, sub-Doppler methods such as pulsed Raman cooling produce nonthermal population distributions that significantly bias measurements of $\bar{n}$ potentially confusing interpretations of heating rate diagnostics and mislead error models \cite{chen2017sympathetic, rasmusson2021optimized, reed2024comparison}. Cooling to near the ground state with only Doppler cooling avoids this complication, enabling faster characterization of the motional distribution and more accurate modeling and accounting of thermal noise.

Finally, with the much lower $\bar{n}$ limit achievable in the resolved Doppler regime, it is natural to ask if other mechanisms begin to surface that limit to the final steady state $\bar{n}_s$. In the next section, we outline how this is indeed the case by expanding the cooling limit derivation to include the larger picture of \textit{any} motional excitation mechanism, not just spontaneous emission (see Fig. \ref{fig:graph}).

\section{Motional state coherence}
\label{sec:motional-state-coherence}
The quantum coherence of trapped-ion motional states is integral to precision measurement and quantum information science operations as motional states mediate rich interactions between trapped-ion internal states \cite{molmer1999multiparticle, schmidt2005spectroscopy, monroe2021programmable}. However, the current level of decoherence of motional states, the rate of interactions mediated by motion, and the remaining thermal spread of cooled harmonic states is limiting. These noise processes limit atomic clock accuracy \cite{marshall2025high}, electric field sensing sensitivity \cite{ivanov2016high, wu2025wideband}, preparation fidelity of highly nonclassical bosonic states \cite{fluhmann2019encoding, matsos2024robust} and entanglement generation between internal qubit states \cite{hughes2025trapped}. Larger secular frequency $\nu$ will diminish the strength of primary noise mechanisms, reduce recoil heating rates, and reduce thermal averaging noise.

\subsection{Motional heating rate scaling}
\label{sec:motional-heating}
A trapped ion gains excess motional energy over time from perturbations by surrounding noisy electric fields. This process is often called ``anomalous'' heating in the literature. Minimizing the impact of this process is a key experimental constraint for which trap electrode design, surface treatments, and even cryogenic vacuum systems have been pursued \cite{chiaverini2014insensitivity, brownnutt2015ion}. Additionally, each measurement of the internal qubit states, using standard state-dependent fluorescence, induces recoil heating which can be orders of magnitude larger than anomalous heating rates \cite{rasmusson2024measurement}. Here we investigate how the anomalous heating and recoil heating rates will change under different experimental design choices that increase $\nu$.

\subsubsection{Anomalous heating}
The heating rate due to anomalous heating mechanisms is commonly defined in the literature as
\begin{equation}
    \dot{\bar{n}}_{\text{an}} = \frac{e^2}{4m\hbar\nu}S_E(\nu)
    \label{eq:anomalous-heating-rate}
\end{equation}
where $e$ is the charge of the electron and $S_E(\nu)$ is the spectral density of the electric-field noise at the motional frequency $\nu$. We assume the usual dependencies $S_E(\nu)\propto \nu^{-\alpha} r_0^{-\beta} T^{+\gamma}$ with T the temperature of the electrodes and typical exponent values being $\alpha \sim 1$, $\beta \sim 4$, and $\gamma \sim 1$ \cite{brownnutt2015ion}. Propagating these definitions to the anomalous heating term, we find the approximate scalings
\begin{equation}
    \dot{\bar{n}}_{\text{an}} \propto \frac{e^2T}{m\nu^2 r_0^4}
    \label{eq:anomalous-heating-rate2}
\end{equation}

At first glance, there seems to be a $1/\nu^2$ reduction in $\dot{\bar{n}}_{\text{an}}$ for increasing $\nu$. First, larger $\nu$ leads to a smaller dipole moment of the oscillating trapped-ion---the prefactor of Eq. \eqref{eq:anomalous-heating-rate}---hence reducing the ion's coupling to electrical field noise. Second, $S_E(\nu)$ will be smaller at higher $\nu$ due to the inverse scaling $S_E(\nu) \propto 1/\nu$. Overall, a potential scaling of $\bar{n}_{\text{an}} \propto 1/\nu^2$.

However, how an increase in $\nu$ is experimentally realized gives different scalings---as was observed in the previous section for cooling rates. Consider varying the charge to mass ratio $Q/m$. Lowering $m$ to reach higher $\nu$ would partially cancel terms that make up $\dot{\bar{n}}_{\text{an}}$ as $\dot{\bar{n}}_{\text{an}} \propto 1/(m\nu^2)$. As the lighter ion couples more strongly to the confining fields, it also is more easily ``pushed'' by the noisy fields. Yet, what frequency the ion is sensitive to in the electric-field noise spectrum will go up, leading to a decrease in power of the noise---assuming $1/f$ like noise. Overall, the heating rate scaling would be $\dot{\bar{n}}_{\text{an}} \propto 1/\nu$.

Varying the electrode distance also has nontrivial consequences. The motional frequency depends on the ion-to-electrode distance $\nu \propto 1/r_0^2$ and the heating rate goes as $\dot{\bar{n}}_{\text{an}} \propto 1/(\nu^2r_0^4)$. Therefore, keeping all other experimental parameters fixed, a change in $r_0$ does not change the heating rate $\dot{\bar{n}}_{\text{an}} \propto 1$! As the ion gets closer to the electrode, the noisy electric field strength increases, increasing the heating rate. However, at a coincidentally equal and opposite rate, the ion is confined to higher frequencies reducing the effective dipole moment and increasing what frequency the ion is sensitive to and therefore lower power noise.

Increasing the confining voltage $V_0$ and $\Omega_{\text{rf}}$ reduces $\dot{\bar{n}}_{\text{an}}$ by increasing, and only affecting, $\nu$. However, new or secondary effects may become more important. For example, at increased RF voltage and frequency, electrodes may heat more which would increase the ion heating rate due to the temperature dependence of $\dot{\bar{n}}_{\text{an}}$.

On the whole, going to higher $\nu$ will reduce the anomalous heating rate, but the experimental realization can affect the precise exponent by which $\dot{\bar{n}}_{\text{an}}$ is reduced.

\begin{figure}
    \centering
    \begin{tikzpicture}[node distance={25mm}, thick, main/.style = {draw, circle}] 
    \node[main] (1) {$n-1$};
    \node[main] (2) [right of=1] {$\textcolor{white}{+}n\textcolor{white}{+}$};
    \node[main] (3) [right of=2] {$n+1$};
    \draw[->] (1) to [out=300, in=240, looseness=1] node[midway, below] {$nA_+$} (2);
    \draw[->] (2) to [out=300, in=240, looseness=1] node[midway, below] {$(n+1)A_+$} (3);
    \draw[->] (3) to [out=120, in=60, looseness=1] node[midway, above] {$(n+1)A_-$} (2);
    \draw[->] (2) to [out=120, in=60, looseness=1] node[midway, above] {$nA_-$} (1);
    \draw[->, bostonuniversityred] (1) -- node[midway, above] {\textcolor{bostonuniversityred}{$\dot{\bar{n}}_{\text{an}}$}} (2);
    \draw[->, bostonuniversityred] (2) -- node[midway, above] {\textcolor{bostonuniversityred}{$\dot{\bar{n}}_{\text{an}}$}} (3);
    \end{tikzpicture}
    \caption{Cooling and heating processes effecting $\bar{n}$ under laser cooling \textit{and} anomalous heating. Operator $A_+$ increases motional occupation while $A_-$ decreases it under the absorption and emission of photons. Anomalous heating, in red, increases motional occupation at a rate $\dot{\bar{n}}_{\text{an}}$. The combined picture of atom-photon processes and ambient noise affecting the motional state changes the commonly derived steady-state mean occupation $\bar{n}_s$ [Eq.~\eqref{eq:steady-state-nbar}] to become a more encompassing model.}
    \label{fig:graph}
\end{figure}
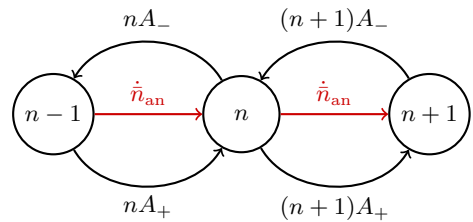

\subsubsection{Cooling limit with anomalous heating rate}

Anomalous heating is a primary heating mechanism during the execution of quantum operations. However, it has not previously been included in the derivation of the cooling limit, which focuses on optical processes. We now broaden the previous cooling limit derivation to include anomalous heating as shown in Fig.~\ref{fig:graph}. The cooling rate, including anomalous heating rate $\dot{\bar{n}}_{\text{an}}$, becomes
\begin{equation}
    \frac{d}{dt}\bar{n} = -(A_- - A_+)\bar{n} + A_+ + \dot{\bar{n}}_{\text{an}} \;.
\end{equation}
The leading terms of the steady state limit are then
\begin{align}
    \bar{n}_s^\prime &= (A_+ +\dot{\bar{n}}_{\text{an}})/(A_- - A_+) \nonumber \\
    & \approx \left(\frac{\Gamma}{2\nu}\right)^2 + \frac{\dot{\bar{n}}_{\text{an}}}{\eta^2 R_{\text{sat}}} \;.
    \label{eq:steady-state-nbar}
\end{align}

The cooling limit now has two contributions as shown in Fig. \ref{fig:cooling-rates-and-limits}(c). First, from the previous mechanism spontaneous emission which is reduced by $\sim 1/\nu^2$ for increasing $\nu$. Second, anomalous heating from the noisy electric field environment which is reduced roughly by $\sim1/\nu$. In Fig. \ref{fig:cooling-rates-and-limits}, we see these two sources can be comparable in the resolved regime ($\nu \gg \Gamma$) with the anomalous heating contribution eventually exceeding spontaneous emission. The exact secular frequency at which this happens highly depends on experimental realization and parameters, but will eventually happen due to the slower scaling of anomalous heating compared to spontaneous emission.

\subsubsection{Recoil (measurement) heating}
Recoil heating during state-dependent fluorescence measurements can easily constitute the largest contribution to motional excitation, even for short detection times \cite{rasmusson2024measurement}. This is particularly salient when considering the requirements for the many mid-circuit measurements in quantum error correction protocols. Most of the laser cooling time, in these protocols, will cool the heating induced by measurement. However, recoil heating is suppressed for large $\nu$.

For state-dependent fluorescence detection ($\Delta = 0$), the $A_-$ coefficient remains the same, and $A_+ = A_-$. The change in the average motional state due to recoil heating during measurement is then given by
\begin{equation}
    \dot{\bar{n}}_{\text{m}} = A_+
\end{equation}
In the resolved sideband regime ($\nu \gg \Gamma$), this evaluates to
\begin{equation}
    \dot{\bar{n}}_{\text{m}} = \eta^2 R_{\text{sc}}\left[1 + \left(\frac{\Gamma}{2\nu}\right)^2\right]
\end{equation}
to second order in $\Gamma/\nu$. The Lamb-Dicke parameter scaling $\eta^2 \propto 1/\nu$ gives a reduction in recoil heating for increasing $\nu$ with the leading term evaluating to
\begin{equation}
    \dot{\bar{n}} \propto \frac{k^2}{m \nu}
    \label{eq:recoil-heating}
\end{equation}
where $k$ is the wavenumber of the incident laser.

A single state-dependent fluorescence measurement can increase $\bar{n}$ as much as 11 quanta as observed in Ref. \cite{rasmusson2024measurement} for a heavy ion at a low secular frequency of $\sim$ 1 MHz. Even a relatively high anomalous heating rate of 1000 q/s over a 2 ms experiment would only heat an ion on average 2 quanta per shot making the recoil heating due to measurement by far the dominant source of motional excitation per cycle. Increasing $\nu$ to $2\pi \times$ 10 MHz (not yet in the resolved regime) could reduce recoil heating to just $\sim 1$ quantum.

\begin{figure}
    \centering
    \includegraphics[width=1\linewidth]{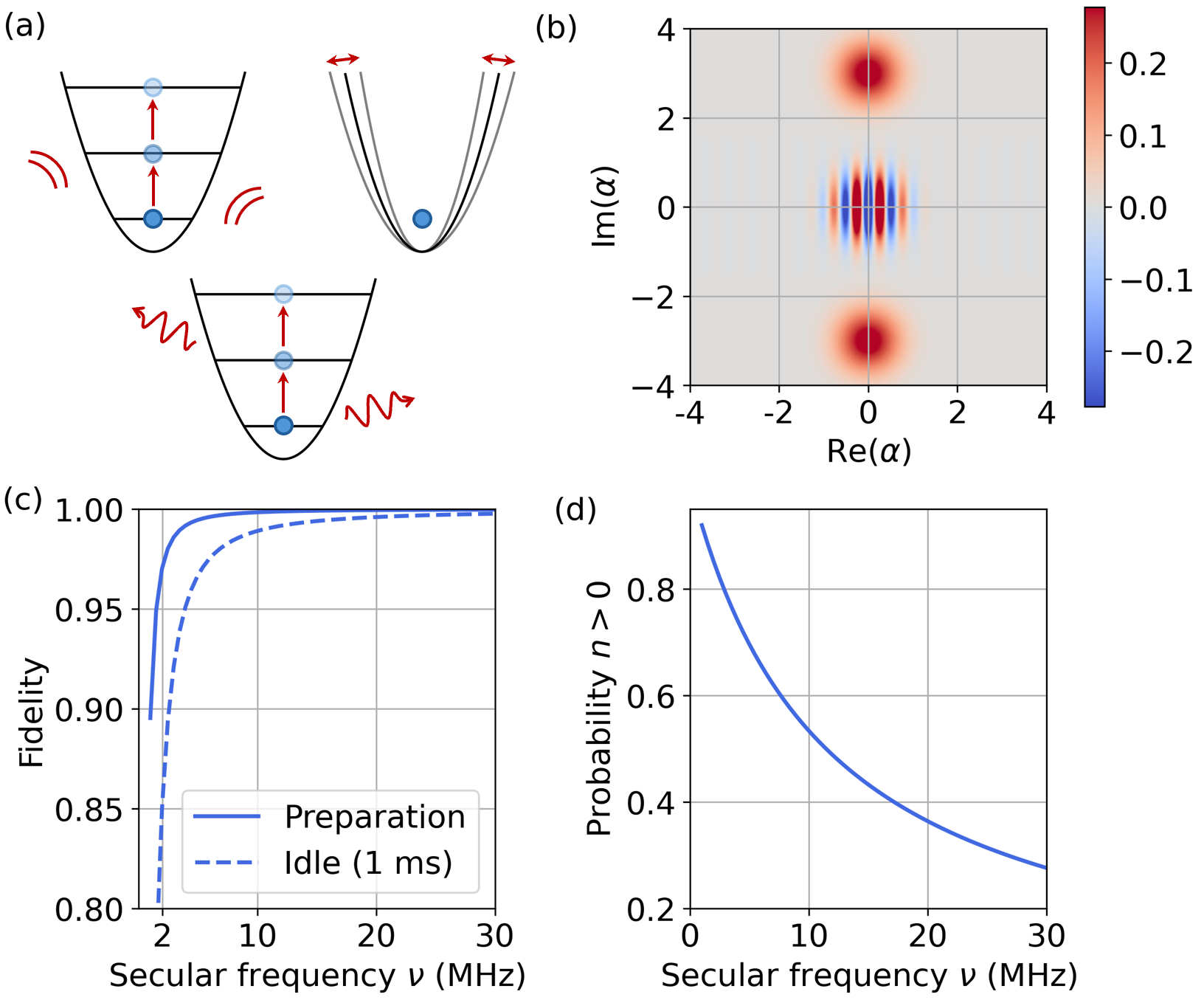}
    \caption{Reduction of primary decoherence mechanisms of trapped-ion quantum motional states for increasing $\nu$. (a) Schematic of decoherence mechanisms for trapped-ion motional states: anomalous heating and dephasing from noisy electric fields and photon recoil heating. (b) Wigner function of the nonclassical ``cat'' state $|i\alpha\rangle + |-i\alpha\rangle$ for $\alpha = 3$ where $|\alpha\rangle = e^{-|\alpha|^2/2}\sum_{n=0}^{\infty}(\alpha^n/\sqrt{n!})|n\rangle$. (c) Preparation fidelity of the $\alpha=3$ cat state assuming a 10 s$^{-1}$ dephasing rate, 100 $s^{-1}$ heating rate, and thermal spread of the initial motional state $\bar{n} = 0.1$ for a $\nu = 2\pi\times 1$ MHz trap. Each error term reduces for increased $\nu$ as described in the text. (d) The probability of exciting the motional ground state $|n=0\rangle$ due to recoil heating. For $\nu = 2\pi\times 2$ MHz, the probability is 85\% and reduces to 25\% for $\nu = 2\pi\times 30$ MHz.}
    \label{fig:cat-state-fid}
\end{figure}

\subsection{Motional dephasing}
\label{sec:motional-dephasing}
Beyond resonant electric-field noise, off-resonant electric-field noise is also a primary decoherence mechanism. It perturbs the trapping curvature $\nu$ and dephases the motional oscillator. This dephasing of motional states limits, for example, the fidelity of preparing bosonic quantum error correction states \cite{fluhmann2019encoding, matsos2024robust, matsos2024universal}. How dephasing rates would change as a function of secular frequency $\nu$ has had little experimental observation \cite{talukdar2016implications}. 

\begin{figure*}[t]
    \centering
    \includegraphics[width=0.89\textwidth]{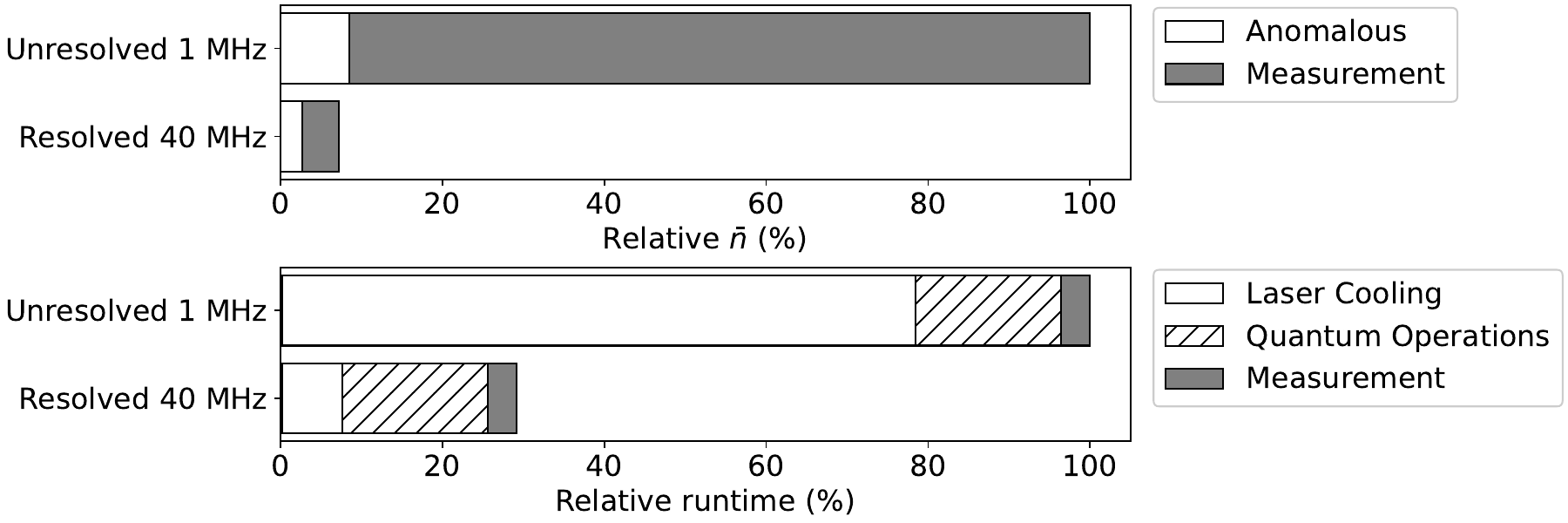}
    \caption{Relative comparison of cooling load and average experimental runtime between an ion trap with 1 MHz or 40 MHz secular frequency. Suppose a 1000 q/s anomalous heating rate, $\Gamma = 2\pi \times 20$ MHz, and target $\bar{n} = 0.15$. (a) Cooling load of the resolved regime relative to the unresolved regime. In the resolved regime, reduction of recoil heating and less time spent cooling reduce the cooling load by more than a factor of 10. (b) Average runtime of the resolved regime relative to the unresolved regime. The reduced cooling load and faster cooling rate compound to lower the time spent laser cooling by a factor of 10.}
    \label{fig:duty-cycle}
\end{figure*}

Unlike anomalous heating noise mechanisms, dephasing noise has off-resonant broadband frequency contributions that change roles when $\nu$ is varied. In general, noise slower than an experimental operation, such as a two-qubit gate, can largely be mitigated by frequent calibration \cite{fluhmann2019encoding} or phase space modulation techniques \cite{valahu2022quantum} as control pulses can be realized quickly enough to counteract these slower noise rates. Noise faster than the experiment drive time ($f_{\text{noise}} \gg \nu$) occurs too quickly to be mitigated, but also constitutes a much smaller portion of the noise power spectral density assuming a $\sim1/f$ like dependence \cite{brownnutt2015ion, talukdar2016implications}. By increasing $\nu$, a larger noise bandwidth becomes slow relative to the ion motion and hence mitigable leaving an approximate $1/\nu$ reduction in dephasing noise for increasing $\nu$.

Suppose a 20 Hz dephasing rate for a 2 MHz secular frequency ion trap, an over estimate from experimentally observed rates \cite{fluhmann2019encoding, matsos2024robust}. Increasing $\nu$ up to 50 MHz would lower dephasing rates to 0.8 Hz, enabling motional state coherence times of longer than 1 second.

Overall, heating and dephasing decoherence mechanisms find significant decrease at high $\nu$. For example, Fig. \ref{fig:cat-state-fid}(c) shows the preparation fidelity of a ``cat'' state---a nonclassical bosonic state prepared in the motional states of a trapped ion \cite{monroe1996schrodinger}---and the fidelity of remaining in that state after 1 ms of idling under the previously discussed decoherence mechanisms. At the secular frequencies of current traps $\nu \sim 2\pi\times1$ MHz, state preparation fidelity is $89.5\%$ for the assumed decoherence rates (see Fig. \ref{fig:cat-state-fid}). For $\nu = 2\pi\times 10$ MHz and $2\pi\times 30$ MHz, state preparation fidelity is $>99.8$~\% and $>99.9$~\% respectively. Fidelity of maintaining the cat state after 1 ms of idling also rapidly increases from $64\%$ at $\nu = 2\pi\times1$ MHz to $99$\% for $\nu = 2\pi\times10$ MHz.

Finally, reaching higher motional frequencies would provide fresh insights into the anomalies of heating and dephasing rates. For example, motional excitation rates have a frequency dependence of $1/\nu^{\alpha}$ with a wide range of observed values from $\alpha = $[-1,6] \cite{brownnutt2015ion}. The highest frequencies explored, to the author's knowledge, is one study reaching $\sim$20 MHz but no further \cite{turchette2000heating}. Observing heating and dephasing rates at much higher frequencies would be revealing.

\section{Runtime Scalability}
\label{sec:fast-duty-cycles}
The scalability of experimental runtimes is a key consideration as system sizes and protocol complexity continue to increase \cite{ransford2025helios}. For example, the runtime of large scale trapped-ion quantum computations can have increased idle time which then can become a larger source of error than two-qubit gates \cite{yamamoto2025quantum}. Additionally, trapped-ion experiments are relatively slow in comparison to other quantum information science platforms with duty cycles times on order $\sim$10 ms per quantum circuit layer \cite{ransford2025helios}. Lowering the duration of the slowest operations, laser cooling and transport, would enable fast quantum feedback, continuous error-corrected quantum operations, and longer and increasingly sophisticated quantum circuits or experiments. We investigate trapped-ion operational runtimes at high $\nu$ and how this impacts the scalability of operations.

\subsection{Fast Duty Cycles}
\label{sec:runtime-scalability}

The majority of runtime in trapped-ion experiments, such as quantum simulations and quantum computations, is spent laser cooling. As much as 69\% of the total runtime in some cases \cite{pino2021demonstration}. To investigate the effect higher $\nu$ will have on the average runtime of an experiment, we define the average shot time to consist of roughly four parts: laser cooling to near the motional ground state $\tau_{\text{lc}}$, qubit state preparation $\tau_{\text{sp}}$, executing the quantum operations of interest $\tau_{\text{op}}$, and measurement $\tau_{\text{m}}$:
\begin{equation}
    \tau_{\text{shot}} = \tau_{\text{lc}} + \tau_{\text{sp}} + \tau_{\text{qo}} + \tau_{\text{m}} \;.
\end{equation}
This may not fit all experiments, but it is general enough for most. For the quantum charge-coupled device (QCCD) ion trap architecture which performs physical movement of the ions \cite{ransford2025helios}, an additional time $\tau_{\text{t}}$ for transport can be included. We neglect it as it will only strength the following arguments.

Ideally, the quantum computation, simulation, sensing, or networking experiment would take the majority of the runtime implying $\tau_{\text{total}} \sim \tau_{\text{qo}}$. However, trapped-ion experiments typically operate in the following approximate ranges: $\tau_{\text{sp}} \sim 10$ $\mu$s, $\tau_{\text{m}} \sim $ $10^2$ $\mu$s, $\tau_{\text{qp}} \sim$ $10^3$ $\mu$s, and $\tau_{\text{lc}} \sim$ $10^4$ $\mu$s.

We approximate the laser cooling time $\tau_{\text{lc}}$ as the average cooling load $\bar{n}_{\text{cl}}$---the average motional excitation accumulated over a shot---divided by the cooling rate $R_{\text{c}}$
\begin{equation}
    \tau_{\text{lc}} = \bar{n}_{\text{cl}} / R_{\text{c}} \;.
    \label{eq:tau-laser-cooling}
\end{equation}
The average cooling load is largely dominated by measurement recoil heating from the end of the previous shot $\dot{\bar{n}}_{\text{m}} \tau_{\text{m}}$ followed by a contribution from anomalous heating over the course of the total time of the previous shot $\dot{\bar{n}}_{\text{a}} \tau_{\text{total}}$. The laser cooling time is then written as
\begin{equation}
    \tau_{\text{lc}} = \left(\dot{\bar{n}}_{\text{m}} \tau_{\text{m}} + \dot{\bar{n}}_{\text{a}} \tau_{\text{total}}\right) / R_{\text{c}} \;.
\end{equation}

\begin{table}[b]
    \caption{Scaling of the cooling time in the unresolved and resolved sideband regimes. Cooling time is split into time spent cooling motional excitations from measurement recoil heating and anomalous heating. The unresolved regime has an additional $\Gamma/\nu$ factor compared to the resolved regime which lengthens the cooling time.}
    \label{tab:duty-cycle-scalings}
    \def\arraystretch{1.5}
    \begin{tabular}{p{3.8cm}|p{1.8cm}|p{1.4cm}} 
     & Unresolved ($\Gamma \gg \nu$) & Resolved ($\nu \gg \Gamma$) \\
     \hline
     recoil ($\dot{\bar{n}}_m \tau_m / R_{\text{c}}$) & $\tau_m \Gamma / \nu$ & $\tau_m$ \\
     \hline
     anomalous ($\dot{\bar{n}}_a \tau_{\text{total}}/ R_{\text{c}}$) & $ \Gamma/\nu^2$ & $1/\nu$ \\
    \end{tabular}
\end{table}

As discussed in the precious sections, the anomalous heating contribution, at best, can scale as $\dot{\bar{n}}_{\text{a}} \propto 1/\nu^2$. Recoil heating scales as $\dot{\bar{n}}_{\text{m}} \propto 1/\nu$.  Finally, the cooling rate scales as $R_{\text{c}} \propto 1/\nu$ in the resolved sideband regime. We can now find the laser cooling time dependency on the relevant experimental design parameters [see Eqs. (\ref{eq:anomalous-heating-rate2}, \ref{eq:recoil-heating})]
\begin{equation}
    \tau_{\text{lc}} \propto \tau_{\text{m}} k^2 + \tau_{\text{total}}/\nu \;.
\end{equation}
In the \textit{resolved} sideband regime, the laser cooling time spent cooling measurement photon recoil heating has no direct motional frequency dependence and is instead largely dependent on the wavenumber of the cooling transition $k$ and the measurement time $\tau_{\text{m}}$. The lack of dependence on $\nu$ makes intuitive sense as the reduced motional coupling at high $\nu$ for the resonant recoil scattering also leads to slower scattering rates for laser cooling. Yet, this is in fact a speed up compared to the unresolved regime of most ion traps which have an addition factor $\Gamma/\nu \sim 10$ (see Table~\ref{tab:duty-cycle-scalings}). Increasing $\nu$ into the resolved sideband regime therefore can give an order-of-magnitude reduction in laser cooling time as shown in Fig. \ref{fig:duty-cycle}. Moving into the resolved sideband regime puts the laser cooling time $\tau_{\text{lc}}$ on par with, or below, typical quantum operations $\tau_{\text{qo}}$ (see Fig. \ref{fig:duty-cycle}).

Additional speed increases could then come from reduced measurement times or cooling transitions with longer wavelengths. The anomalous heating contribution is much smaller, and will continue to diminish for increasing $1/\nu$.

\subsubsection{Quantum Error Correction Cooling Load}
By significantly reducing cooling times, larger scale and more complex experimental directions are made attainable. Consider quantum error correction (QEC) protocols which utilize mid-circuit measurements to extract error information. QEC protocols will have much higher cooling loads because of the increased number of measurements per shot and thus lead to much slower runtimes limiting the rate at which errors can be corrected. This cascades into increased idling and memory errors which are usually not limiting but have been observed to become the dominate noise contribution (larger than two-qubit gate errors) in quantum circuits with QEC protocols \cite{yamamoto2025quantum}.

Consider a generic quantum stabilizer code $[[n,k,d]]$ consisting of $n$ physical qubits, $k$ logical qubits, and code distance $d$. In such a code, there will be a minimum of $n-k$ independent syndrome measurements per error correction round. Each syndrome measurement induces recoil heating. Accounting for $r$ rounds of QEC each with $n-k$ mid-circuit measurements, the average motional excitation gain per experimental cycle has a lower bound of
\begin{equation}
    \bar{n}_{\text{QEC}} \geq r (n-k) \bar{n}_m \;
\end{equation}
which will need to be cooled mid-circuit to maintain high-fidelity operation.  

\begin{figure}[t]
    \centering
    \includegraphics[width=0.8\linewidth]{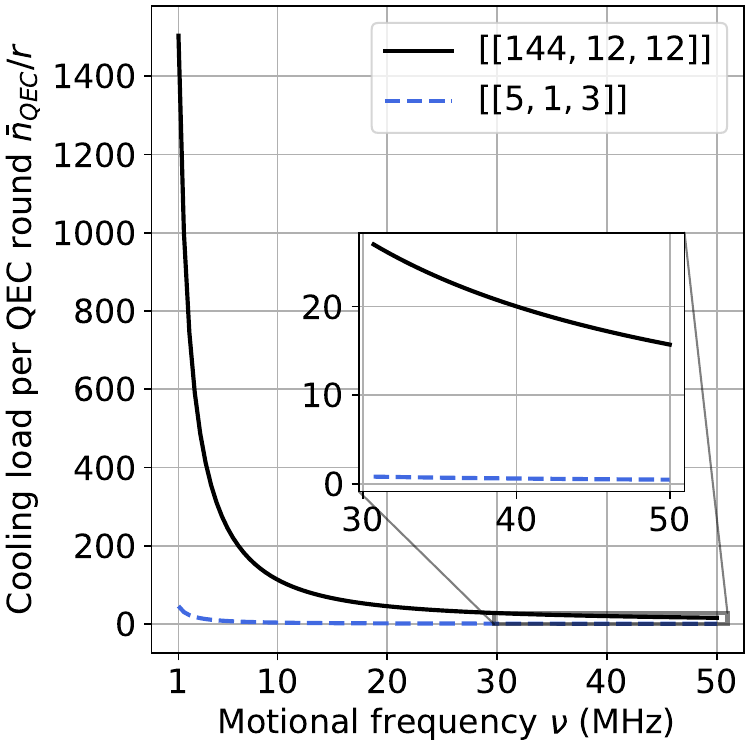}
    \caption{Cooling load $\bar{n}_{\text{QEC}}$ per QEC round $r$. Comparison between the smallest QEC code $[[5,1,3]]$ and a larger $[[144,12,12]]$ code. Initial cooling loads at $\nu = 2\pi\times 1$ Mhz are $\bar{n}_{\text{QEC}}/r=44$ and  $\bar{n}_{\text{QEC}}/r=1452$ for codes $[[5,1,3]]$ and $[[144,12,12]]$, respectively. The cooling load drops off quickly for both codes once $\nu$ is beyond the low frequency regime. Near an intermediate motional frequency of $\nu \sim 2\pi\times30$ MHz, the cooling loads per QEC round $r$ for the codes $[[5,1,3]]$ and $[[144,12,12]]$ are $\bar{n}_{\text{QEC}}/r=0.8$ and $\bar{n}_{\text{QEC}}/r=27$ respectively.}
    \label{fig:qec-cooling-load}
\end{figure}

Suppose a single measurement induces a cooling load of $\bar{n}_m = 11$, a slight underestimate from observations in Ref \cite{rasmusson2024measurement}. For the smallest stabilizer code $[[5,1,3]]$, assuming a minimum number of syndrome checks, the average cooling load per QEC round $r$ is $\bar{n}_{\text{cl}}/r = 44$ quanta (see Fig. \ref{fig:qec-cooling-load}). For the larger $[[144, 12, 12]]$ bivariate bicycle code proposed for larger scale systems \cite{yoder2025tour, ye2025beam}, again with the minimum number of checks, the average cooling load per QEC round would be $\bar{n}_{\text{cl}}/r = 1452$ quanta. These cooling loads present a significant challenge to the scalability of operations. Yet, it should be noted that if all ions contribute to cooling, these cooling loads effectively reduce just to $\bar{n}_{\text{cl}}/r = 10$ as each ion that heated could also cool. However, there are significant practical challenges to this possibility. First, spontaneous emission during cooling would destroy encoded information in the internal or motional states, so cooling must be left to auxiliary ions, ``coolant'' ions, that do not have any quantum information stored. Second, all coolant ions may not be able to be cooled to the ground state simultaneously, especially for large scale numbers of ions, due to laser related constraints. Cooling to near the motional ground state in low frequency traps requires pulsed Raman sideband cooling techniques, which require high power and well-controlled laser systems and would likely not be able to address all coolant ions simultaneously. Overall, there is a significant QEC cooling load that could be mitigated but still needs core challenges addressed.

A high motional frequency ion trap directly alleviates these difficulties. In a moderately high frequency trap with $\nu = 2\pi \times 30$ MHz, the $[[144, 12, 12]]$ bivariate bicycle code cooling load per QEC round reduces to just $\bar{n}_{\text{cl}} = 28$ quanta as shown in Fig. \ref{fig:qec-cooling-load}. In the high frequency regime, Doppler cooling to near the motional ground state requires simpler laser systems and lower powers enabling more scalable optical systems such as sheet beams or integrated photonics delivery cooling light to more ions \cite{moses2023race, kwon2024multi}. Additionally, if reaching high frequency is technically challenging across the entire QCCD trap, dedicated zones could be designed for high motional frequency.

\subsection{Transport speed}
\label{sec:transport-speed}
In the QCCD architecture, trapped ions are physically transported enabling scalability while maintaining all-to-all connectivity \cite{moses2023race, ransford2025helios}. Transport protocols that move, merge, and split ions are essential, but can take a significant fraction of runtime and are often executed in the adiabatic regime to avoid motional excitation and maintain scalable transport waveforms \cite{pino2021demonstration, moses2023race, ransford2025helios}.

The adiabatic condition for ramping a harmonic well is given by $\frac{1}{\nu^2}\frac{dt}{d\nu} \ll 1$ \cite{chen2010fast, bowler2012coherent}. Therefore, increasing $\nu$ by an order of magnitude, enables much faster transport speeds while maintaining adiabaticity and thus low motional excitation. Suppose transport protocols for a quantum computation takes 20 ms with a center-of-mass motional mode of 1 MHz. At a high frequency of 50 MHz, the same level of adiabaticity of the original transport protocol can be realized on the order of only $\sim$10 $\mu$s. Such a dramatic speed up will address one of the primary bottlenecks in scaling trapped-ion QCCD runtimes without suffering additional motional excitation \cite{ransford2025helios}.

\subsection{Gate speed}
\label{sec:gate-speed}

Finally, we consider the effect of high frequency $\nu$ on two-qubit interactions speeds. Two-qubit gates enable universal quantum computation and are a foundational technique for quantum information science \cite{haffner2008quantum}. Two-qubit gates for trapped-ions are commonly realized through the M\o lmer-S\o rensen (MS) interaction which mediates qubit-qubit interactions through the shared motion of co-trapped ions \cite{molmer1999multiparticle}. MS gate operations go slower for larger $\nu$ due to smaller spin-motion coupling $\eta$, but the reduction is relatively weak at $\tau_{\text{g}} \sim \sqrt{\nu}$.

One advantage that may be worth the speed reduction of the MS interaction at large $\nu$ is that spectral crowding of motional modes, for multiple co-trapped ions, is reduced. Off-resonant coupling to unwanted motional modes and the carrier transition is reduced by $\sim 1/\Delta^2$ for detuning $\Delta$ enabling selective mode coupling even with global laser drives. With more selective mode couplings, interaction graphs for quantum simulation experiments can have new connectivity \cite{kyprianidis2024interaction} and multi-mode coupling drives can realize multi-qubit gates for more efficient quantum computation and simulation \cite{solomons2025full, richerme2025multi}.

The motional vibration frequency sets a fundamental speed limit for motion coupling between ions, so it seems possible that increasing motional vibration speeds could decrease motion-mediated gate times. A different technique using optically induced spin-dependent kicks can directly realize such a speed limited regime for two-qubit gates \cite{savill2025high}. Gate times are limited by the speed of motional vibration across the co-trapped ion crystal and therefore have two-qubit gate times that decrease proportionally to increases in the motional frequency $\tau_{\text{g}} \sim 1/\nu$. This gate scheme, opposed to the MS interaction, would directly realize faster quantum operations at large $\nu$.

\section{Conclusion}
\label{sec:conclusion}

In conclusion, the motional frequency of a trapped ion can be increased a factor of 10 or more to realize an underexplored regime of trapped-ion quantum information science. At the hardware level, faster cooling rates, lower cooling limits, duty cycle speed ups, lower cooling load, and longer motional coherence are possible and set the foundation for promising new directions in quantum information science such as preparing high-fidelity and long lived nonclassical bosonic states and scalable operations for quantum error correction.

This work has found that for increasing $\nu$, the laser cooling limit can transition to being dominated by anomalous heating, not spontaneous emission. Higher motional frequencies have a compounding affect on reducing the experimental runtime. The cooling rate is faster and the cooling load is smaller largely due to reductions in recoil heating. Order-of-magnitude improvements in laser cooling time and transport, the two slowest operations in the QCCD architecture, can be realized with high frequency traps and enable scalable operations for increasing sophisticated protocols in quantum information science.

\begin{acknowledgments}
This work is supported by the College of Computational, Mathematical, and Physical Sciences at Brigham Young University. We thank Philip Richerme for early and insightful discussions.
\end{acknowledgments}

\bibliographystyle{prsty}
\bibliography{main}{}

\end{document}